\documentclass[traditabstract]{aa}  
\usepackage{graphicx,amsmath}

\begin{document}
\title{
GMRT search for 150~MHz radio emission from the transiting extrasolar planets HD189733\,b and 
HD209458\,b\thanks{Data for this observations can be retrieved electronically 
on the GMRT archive server {\tt http://ncra.tifr.res.in/\symbol{126}gmrtarchive} 
and by request to {\tt archive@gmrt.ncra.tifr.res.in}.}
    }

   \author{
 A.~Lecavelier des Etangs\inst{1,2}
\and
 S.~K.~Sirothia\inst{3}
  \and
 Gopal-Krishna\inst{3}       
 \and
 P.~Zarka\inst{4}
   }
   
\titlerunning{GMRT search for 150~MHz radio emission from HD189733\,b and HD209458\,b}

\offprints{A.L. (\email{lecaveli@iap.fr})}

   \institute{
   CNRS, UMR 7095, 
   Institut d'Astrophysique de Paris, 
   98$^{\rm bis}$ boulevard Arago, F-75014 Paris, France
   \and
   UPMC Univ. Paris 6, UMR 7095, 
   Institut d'Astrophysique de Paris, 
   98$^{\rm bis}$ boulevard Arago, F-75014 Paris, France
   \and
   National Centre for Radio Astrophysics, TIFR, 
   Post Bag 3, Pune University Campus, Pune 411007, India
   \and
   LESIA, Observatoire de Paris, CNRS, UPMC, Universit\'e Paris Diderot, 
   5 place Jules Janssen, 92190 Meudon, France
}   
   \date{} 
 
  \abstract
{
%

We report a sensitive search for meter-wavelength emission at 150~MHz 
from two prominent transiting extrasolar planets, HD189733b and HD209458b.  
To distinguish any planetary emission from possible
stellar or background contributions, 
we monitored these systems just prior to, during, and after the planet's eclipse behind 
the host star. 
No emission was detected from \object{HD209458\,b} with a 3$\sigma$ upper limit of 3.6~mJy.
For \object{HD189733\,b} we obtain a 3$\sigma$ upper limit of 2.1~mJy and a marginal 2.7$\sigma$ 
detection of $\sim$1900$\pm$700~$\mu$Jy from a direction just 13\arcsec\ from the star's 
coordinates ({\it i.e.}, within the beam), 
but its association with the planet remains unconfirmed. 
Thus, the present GMRT observations
provide unprecedentedly tight upper limits for meter wavelengths emissions from
these nearest two transiting type exoplanets. 
We point out possible explanations of the non-detections and briefly discuss the
resulting constraints on these systems.}

\keywords{Stars: planetary systems - Stars: coronae - Techniques: interferometric}

   \maketitle

\section{Introduction}
\label{Introduction}

Because the radio-frequency emission from planets is expected to be 
strongly influenced by their interaction with the magnetic field and 
corona of the host star, the physics of this interaction can be
effectively constrained by obtaining good measurements of the 
properties of the planetary radio emission. Therefore,  
in parallel with theoretical estimates for radio emission from a large number 
of extrasolar planets ({\it e.g.,} Grie\ss meier et al.\ 2007), 
searches have been undertaken for decameter- and meter-wavelength radio 
emission from a few carefully selected extrasolar planets ({\it e.g.,} 
Bastian et al.\ 2000; Ryabov et al.\ 2004; Winterhalter et al.\ 2005; 
George \& Stevens 2007; Lazio \& Farrell 2007; Smith et al.\ 2009) and,
most recently, by Lecavelier des Etangs et al.\ (2009) at 244 and 610~MHz and
Lazio et al.\ (2010) at 325 and 1425~MHz with
much improved sensitivity. Based on the theoretical estimates,
radio detections seem currently feasible only provided that the planets
are 10$^3$ to 10$^4$ times stronger emitters than Jupiter. However, the 
extreme physical conditions of ``hot-Jupiters'' could make this realizable, 
e.g., through a massive infall of Poynting flux on them (Zarka 2007). This provides
the physical justification for the radio-magnetic scaling law proposed 
by Zarka et al.\ (2001). In any event, it should be borne in mind that all 
existing estimates of cyclotron maser decametric emission are based on a 
host of unknowns, {\it e.g.}, stellar winds, coronal density, and stellar 
and planetary magnetic fields.

Practically all searches for radio emission from exoplanets have been
carried out at metre wavelengths and, to date, the results have been 
negative.
The telescopes used include
UTR (10-30\,MHz, $\sigma$$\sim$1.6\,Jy), VLA (74\,MHz, $\sigma$$\sim$50\,mJy; 
325\,MHz, $\sigma$$\sim$0.58\,mJy; 
1425\,MHz, $\sigma$$\sim$16\,$\mu$Jy), 
and GMRT (150\,MHz, $\sigma$$\sim$10\,mJy; 244\,MHz, $\sigma$$\sim$0.7\,mJy; 
610\,MHz, $\sigma$$\sim$50\,$\mu$Jy; see review in Lazio et al.\ 2009). 
At these low frequencies, the principal contributors to the noise level
are the sky background, radio frequency interference and the ionospheric 
scintillations that distort the incoming signal and increase the noise.
Hence, interferometric observations of high sensitivity and resolution hold 
considerable promise. In particular, the Giant Metrewave 
Radio Telescope (GMRT), a 30-km baseline array consisting of 30 dishes of 45
metre diameter each (Swarup 1990), seems very appealing for this purpose.
In a recent paper (Lecavelier des Etangs et al.\ 2009) we reported the 
first GMRT search, which was targeted at the planet HD189733\,b,  
one of the best candidates among the known ``hot-Jupiter'' type extrasolar planets
(Sect.~\ref{HD189733b}). Our search was more than an order of magnitude more
sensitive than the previously reported deepest search for meter-wavelength emission 
from this system by Smith et al.\ (2009), who observed it at 307-347\,MHz.
For stellar+planetary emission toward HD189733\,b we obtained 3$\sigma$ 
upper limits of 2~mJy at 244~MHz and 160~$\mu$Jy at 610~MHz.
Because one of the plausible explanations for the non-detection is that the
emission frequency is lower owing to a weak planetary magnetic field, 
we decided to make the radio search at 150 MHz, which is the lowest frequency 
accessible using the GMRT. 
The targets chosen are the two most prominent transiting hot-Jupiters: 
\object{HD189733\,b} (Bouchy et al.\ 2005) and \object{HD209458\,b} 
(Mazeh et al.\ 2000; Charbonneau et al.\ 2000).

\section{The targets and the observational strategy}

\subsection{HD189733\,b}
\label{HD189733b}

Located just 19.3~parsec away, HD189733\,b is one of the most prominent 
extrasolar planets known (Bouchy et al.\ 2005). With a semi-major axis 
of 0.03~AU and an orbital period of 2.2 days, it belongs to the class of 
``very hot-Jupiters". More importantly, because this planet is seen to
transit its parent star, the transits and eclipses have been
used to probe the planet's atmosphere and environment 
({\it e.g.}, Charbonneau et al.\ 2008; D\'esert et al.\ 2009).    

HD189733\,b orbits a small and bright main-sequence K-type star and shows a 
transit occultation depth of $\approx$2.5\% at optical wavelengths
(Pont et al.\ 2007). The
planet has a mass $M_p$=1.13~Jupiter mass ($M_{\rm Jup}$) and a
radius $R_p$=1.16~Jupiter radius ($R_{\rm Jup}$) in the visible (Bakos
et al.\ 2006; Winn et al.\ 2007). The orbital period of the planet (2.21858~days) 
and its transit epochs have been measured precisely 
(H\'ebrard \& Lecavelier des Etangs 2006; Knutson et al.\ 2009). 
Sodium absorption has been detected in the planet's atmosphere
by ground-based observations (Redfield et al.\ 2008). 
Using the `Advanced Camera for Surveys' onboard the Hubble Space Telescope 
(HST), Pont et al.\ (2008) detected atmospheric haze, which is interpreted as Mie
scattering by small particles (Lecavelier des Etangs et al.\ 2008a). 
Carbon monoxide molecules have been tentatively invoked to explain the excess 
absorption seen at 4.5~$\mu$m (Charbonneau et al.\ 2008; D\'esert et al.\ 2009). 
Absorption in Lyman-$\alpha$ observed with the HST/ACS is explained in terms
of atomic hydrogen escaping from the planet's exosphere at a rate 
of 10$^7$-10$^{11}$~g/s (Lecavelier des Etangs et al., 2010). 
Recent XMM-Newton observations show flare-type X-ray emission from HD189733 and
a possible softening of the X-ray spectrum during the planetary 
eclipse (Pilliteri et al.\ 2010).  
Spectro-polarimetry has revealed the strength and topology 
of the stellar magnetic field, which reaches up to 40~G (Moutou et al.\ 2007).
Fares et al.\ (2010) found a variable magnetic field on the surface of the 
star from 22 to 36~G. Using a model to extrapolate the magnetic field, the authors 
concluded that the average magnetic field should be in the range 4-23~mG 
at the distance of the planet. 

\subsection{HD209458\,b}
\label{HD209458b}

Located 47~parsecs away, HD209458\,b is the second-closest 
transiting hot-Jupiter after HD189733\,b. 
This planet is the first extrasolar planet
for which atmosphere and exosphere have been probed using the transit
technique (Charbonneau et al.\ 2002; Vidal-Madjar et al.\ 2003).
With a semi-major axis 
of 0.047~AU and an orbital period of 3.5 days (Mazeh et al.\ 2000)
this planet is a prominent member of the "hot-Jupiter" class. 
As for HD189733\,b, the planetary transits and eclipses have 
been used to probe the planet's atmosphere and environment 
({\it e.g.}, Sing et al.\ 2008; Lecavelier des Etangs et al.\ 2008b).    

HD209458\,b orbits a main-sequence G-type star and shows a 
transit occultation depth of $\approx$1.5\% at optical wavelengths
(Sing et al.\ 2008). The
planet has a mass $M_p$=0.69~Jupiter mass ($M_{\rm Jup}$) and a
radius $R_p$=1.32~Jupiter radius ($R_{\rm Jup}$). 
Despite the large body of data available for this system, 
little is known about its magnetospheric activity or 
the star-planet interactions. Nonetheless, a striking
outflow of hydrogen gas has been detected 
(Vidal-Madjar et al.\ 2003, 2008; Ehrenreich et al.\ 2008); 
these atoms are probably expelled at high velocity caused by the 
stellar radiation pressure (Lecavelier des Etangs et al.\ 2008c).
High velocity ions (C\,{\sc ii}, Si\,{\sc iii}), possibly at high temperature, 
have also been detected with HST/STIS and HST/COS (Vidal-Madjar et al. 2004; 
Ben-Jaffel \& Sona Hosseini 2010; Linsky et al.\ 2010).
Possible auroral dayglow emission from the planet's atmosphere was reported
by France et al.\ (2010), but the excitation mechanism for the detected
transition of the H$_2$ molecules remains unclear. 

\subsection{The planetary eclipses at 150MHz}

The atmosphere and environment of transiting planets are
studied using the planetary eclipse technique, that is, by 
subtracting the signal received when the planet 
is hidden behind the star from observations made before and after the
eclipse.
This allows a reliable extraction of the contribution of the planet 
and its atmosphere to the received signal. 
This technique has
made possible the detection of thermal infrared emission 
from an extensive sample of 
extrasolar planets using both Spitzer ({\it e.g.}, 
Charbonneau et al.\ 2005; Deming et al.\ 2005; Stevenson et al.\ 2010) and 
several ground based telescopes ({\it e.g.}, Sing \& L{\'o}pez-Morales 2009; 
Croll et al.\ 2010; C{\'a}ceres et al.\ 2011; de Mooij et al.\ 2011).
Using Spitzer 
spectroscopy of planetary eclipses, infrared spectra of the atmosphere of HD189733\,b 
have revealed signatures of H$_2$O absorption and possibly 
weather-like variations in the atmospheric conditions (Grillmair et al.\ 2009; 
see Seager \& Deming 2010 
for a review on the Spitzer observations of planetary atmospheres).

We have adopted a similar strategy by monitoring the radio flux from 
the \object{HD189733} and \object{HD209458} systems before, 
during, and after the eclipse of the respective planet. For this 
we developed a technique to extract the time series of the target's
radio continuum flux from the interferometric visibilities measured with the GMRT 
(Sect.~\ref{Observations and data analysis}). 
By comparing  the radio flux levels at different phases of the planetary orbit, 
we can thus distinguish between any radio emission 
contributed by the planet and by the host star 
(or any other background source within 
the synthesized beam). 

The choice of 150~MHz, the lowest frequency available at GMRT, was 
intended to increase the likelihood of detection
for a weak planetary magnetic field. Indeed, in our previous search 
using GMRT the non-detection down to the very low limit of
a few hundreds of $\mu$Jy 
is most plausibly explained in terms of a weak planetary magnetic field 
and the low gyrofrequency of the resulting
electron-cyclotron maser radiation. 

\section{Observations and data analysis}
\label{Observations and data analysis}

The GMRT 150~MHz observations of the HD189733b field were made 
on 2009 August 15, and of HD209458b field on 2009 September 9. 
For each observation, the phase centre was set at the star's position :
$\alpha_0$=20h00m43.7s, $\delta_0$=+22\degr42\arcmin39\arcsec (J2000)
for
HD189733, and 
$\alpha_0$=22h03m10.8s, $\delta_0$=+18\degr53\arcmin03\arcsec (J2000) 
for HD209458.

\begin{figure*}[tbh]
\begin{center}
\hbox{
 \includegraphics[angle=0,width=\columnwidth, viewport=30 214 570 620,clip]
 {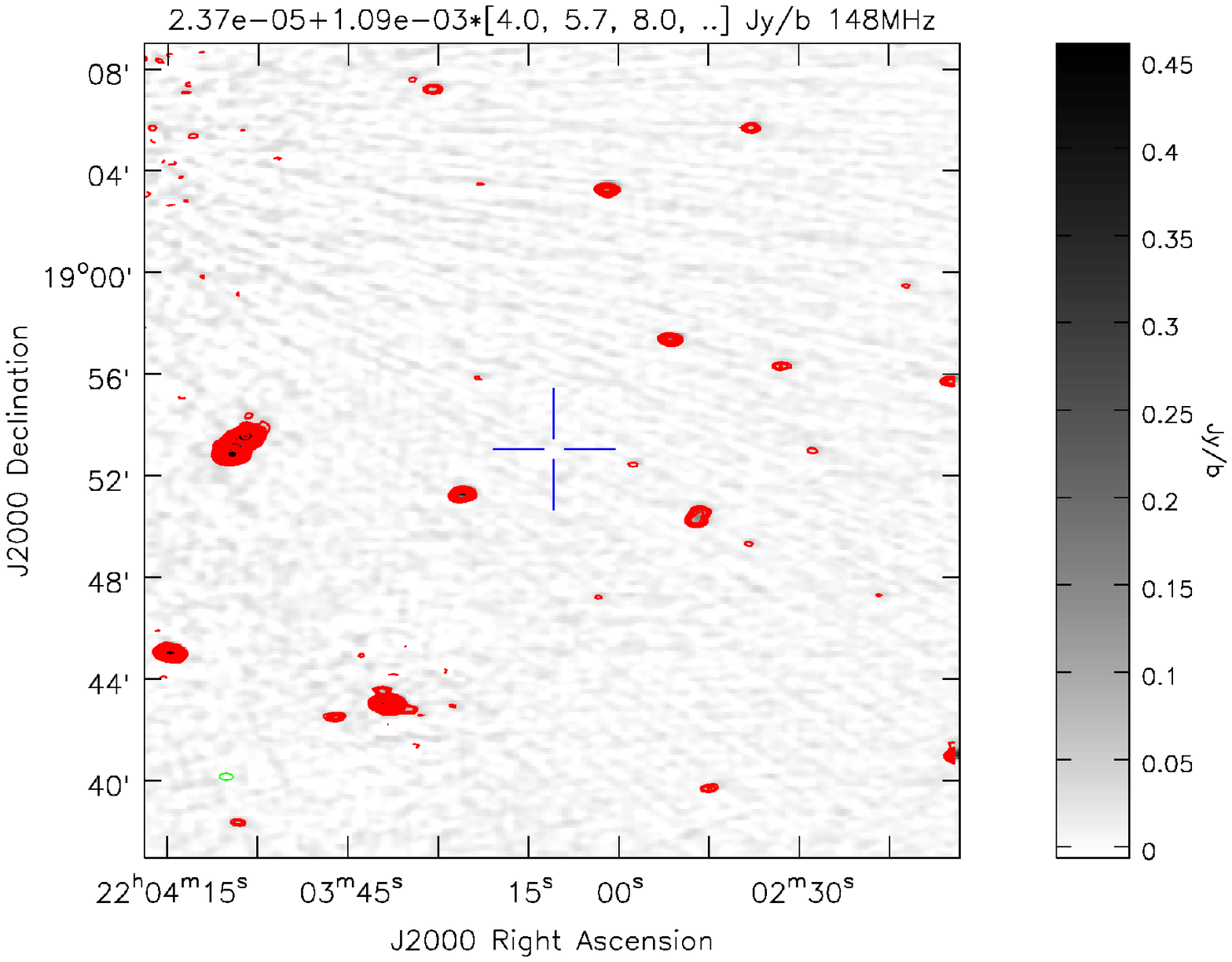}
 \includegraphics[angle=0,width=\columnwidth, viewport=30 214 570 620,clip]
 {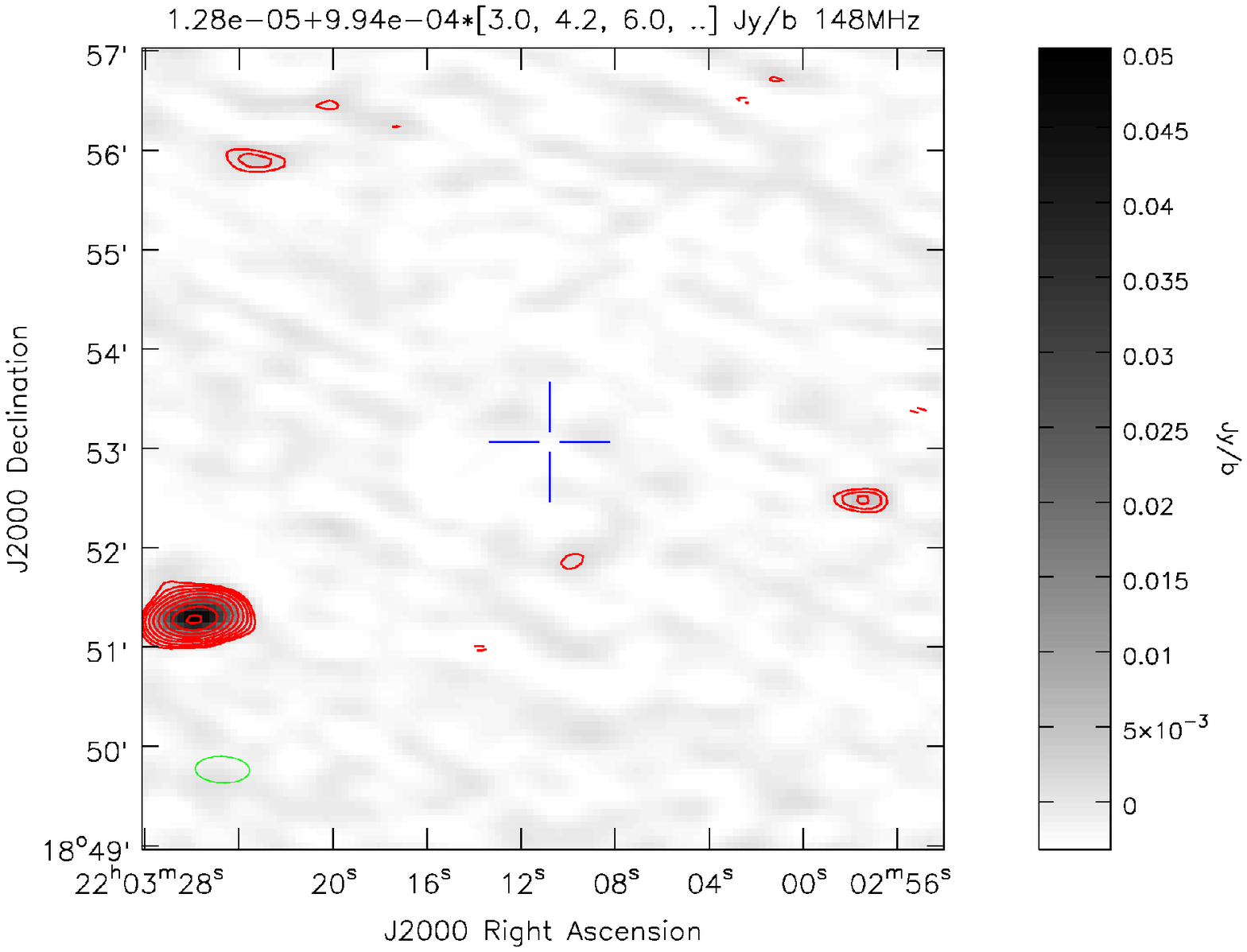}
}
\caption{
GMRT image of HD209458 field at 150 MHz.
The green ellipse in the lower left corner of each image shows the half-power beamwidth 
(33\arcsec$\times$16\arcsec at a position angle of 88$\degr$).
The blue cross indicates the center of field toward HD209458. 
In the left panel the contour levels are at 4.0, 5.7, and 8.0 times the image RMS of 1100\,$\mu$Jy per beam$^{-1}$. 
In the right panel the contour levels are at 3.0, 4.2, and 6.0 times the center field RMS of 990\,$\mu$Jy per beam$^{-1}$. 
Negative contours appear as dashed lines.}
\label{HD209458_map}
\end{center}
\end{figure*}

\begin{figure}[tbh]
\begin{center}
\hbox{
\includegraphics[angle=90,width=\columnwidth]
{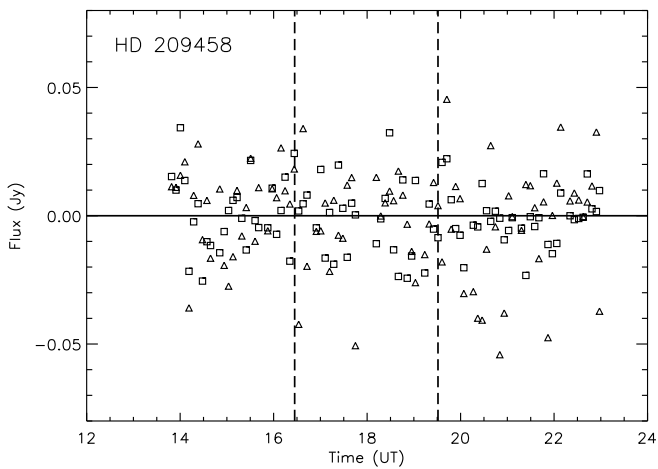}
}
\caption{Time series of the flux density in the direction of HD209458, for a sampling window of
339~seconds. The vertical dotted lines indicate the beginning and the end of the planet's
eclipse behind the host star. Triangles and squares correspond to the RR and LL polarizations, 
respectively.
}
\label{HD209458_ts}
\end{center}
\end{figure}

The HD189733 observations started at 13h30m~UT and finished at 22h30m~UT, 
covering the full passage of the target in the visibility window of the 
GMRT sky and also covering the planet's eclipse behind the star, 
which took place between 
17h27m~UT and 19h16m~UT. 
The HD209458 observations started at 14h30m~UT and finished at 23h30m~UT, 
encompassing the planet's eclipse from 16h27m~UT to 19h31m~UT.

The center frequency of the receiver was set at 148~MHz, with a bandwidth of 16~MHz.
The visibility integration time was 2.09~second.
For each of the two observing runs the total observation time was close to 9~hours, 
including the target field and the calibration sources.

3C48 was observed as the primary flux density 
and bandpass calibrator.
The phase calibrators used are J1924+334 for HD189733 and 
J2251+188 for HD209458; they were observed repeatedly during the observations.
The total data acquisition time on the target field was 7.2~hours for HD189733 and
7.8~hours for HD209458.
 
The data reduction was carried out mainly using the {\tt AIPS++} package 
(version: 1.9, build \#1556).
After applying bandpass 
corrections using 3C48, gain and phase variations of individual antennas 
were quantified and
used for calibrating the flux density, bandpass, gain, and phase for the
target field data. For 3C48 we took a flux density 
of 64.4\,Jy at 150\,MHz (Perley \& Taylor 1999).
The calibrated flux densities toward HD189733 have also been corrected upward 
by a factor 1.6, which is the ratio between the system temperatures 
toward the direction of HD189733 ($T_{\rm sys}$(150MHz)$\approx$1014\,K) 
and the direction of the calibrator 3C48 ($T_{\rm sys}$(150MHz)$\approx$642\,K).

\begin{figure*}[tbh]
\begin{center}
\hbox{
 \includegraphics[angle=0,width=\columnwidth, viewport=30 214 570 620,clip]
 {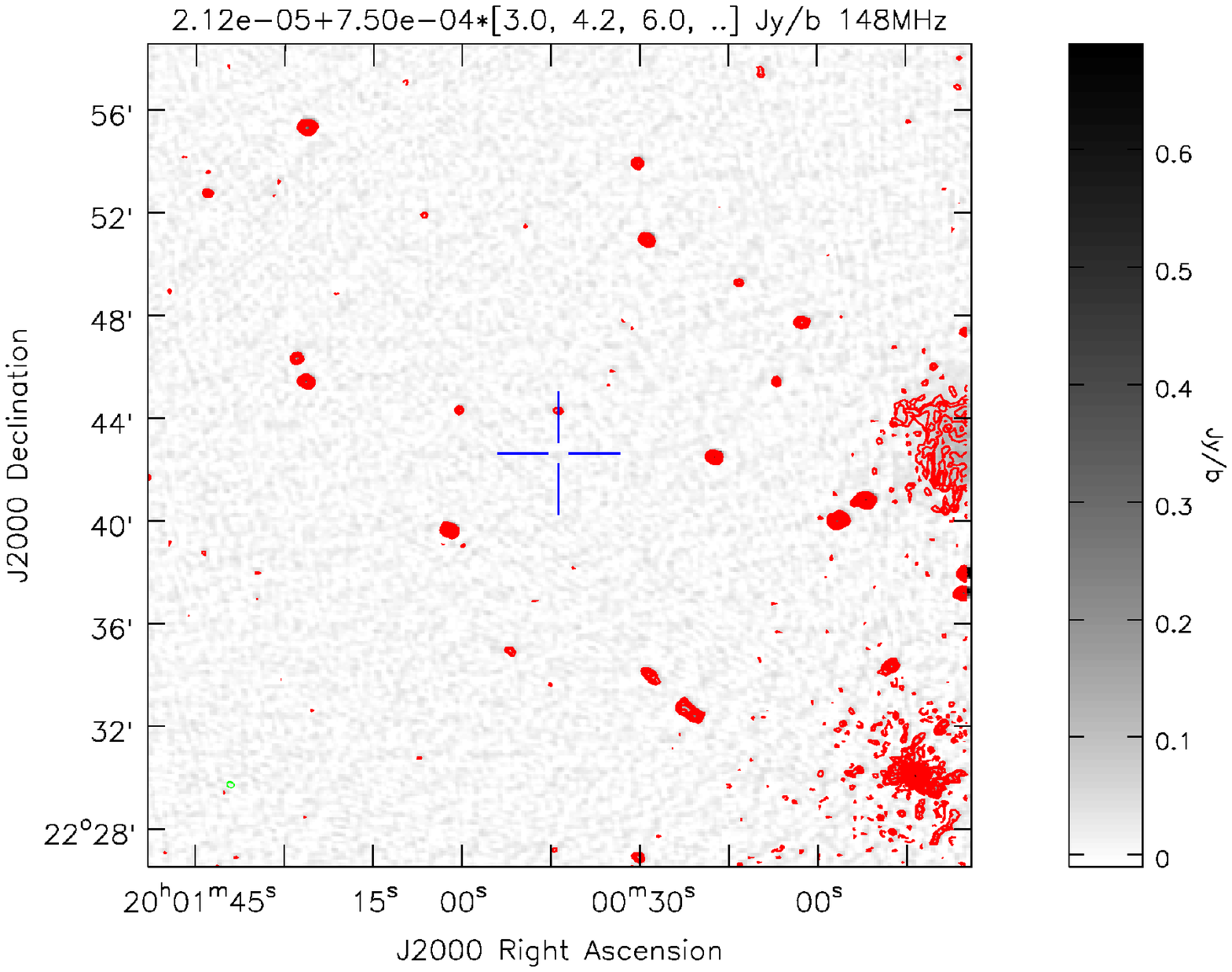}
 \includegraphics[angle=0,width=\columnwidth, viewport=30 214 570 620,clip]
 {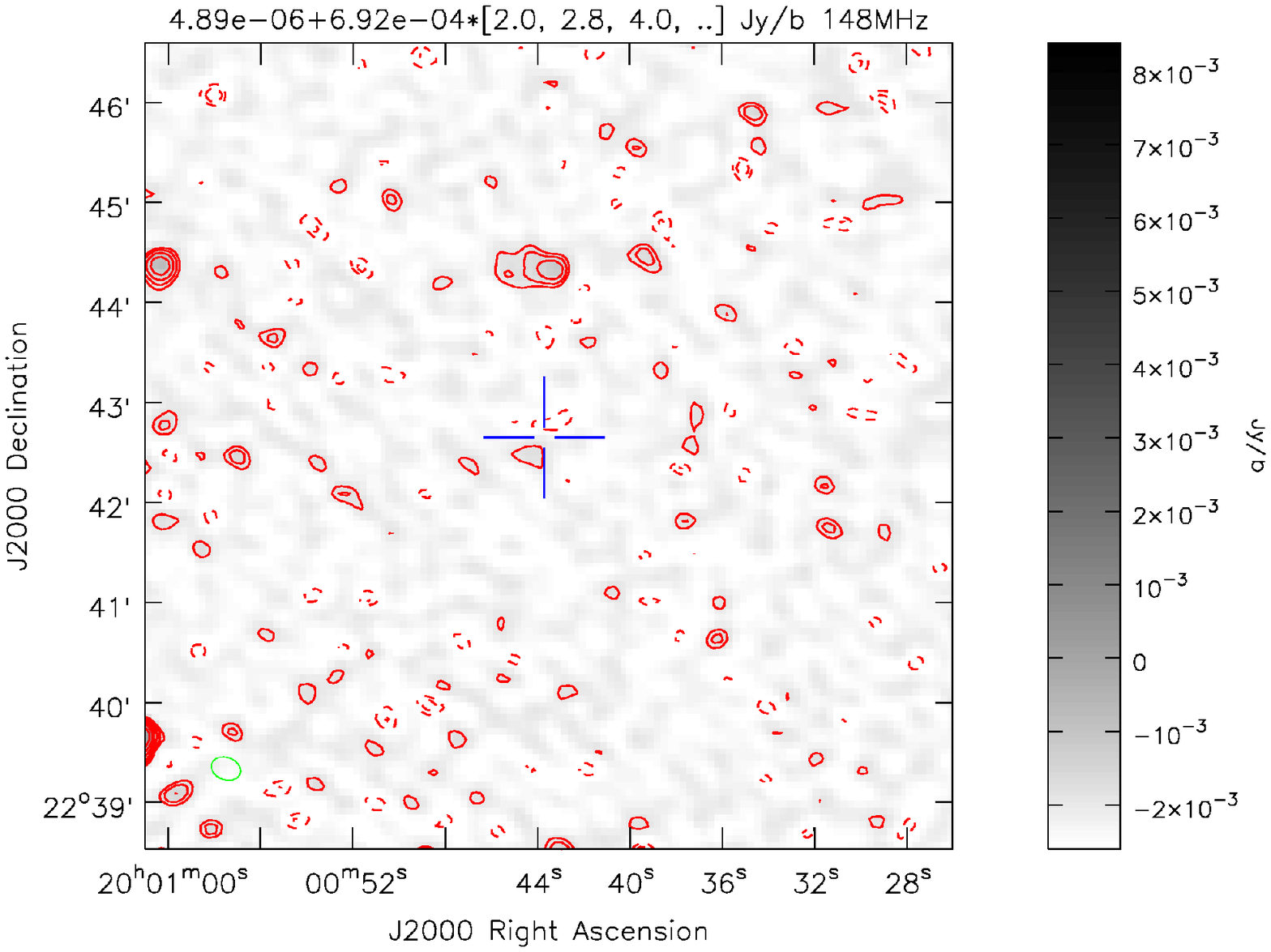}
}
\caption{
Same as Fig.~\ref{HD209458_map} for the HD189733 field at 150 MHz. 
The half-power beamwidth is 18\arcsec$\times$13\arcsec at a position angle of 70$\degr$. 
In the left panel the contour levels are at 3.0, 4.2, and 6.0 times the image RMS of 750\,$\mu$Jy per beam$^{-1}$. 
In the right panel the contour levels are at 2.0, 2.8, and 4.0 times the center field RMS of 690\,$\mu$Jy per beam$^{-1}$, 
showing the 2.7-$\sigma$ source at 13\arcsec\ from HD189733 
to the south\,east of the target position (see text). 
}
\label{HD189733_map}
\end{center}
\end{figure*}

\begin{figure}[tbh]
\begin{center}
\hbox{
\includegraphics[angle=90,width=\columnwidth]
{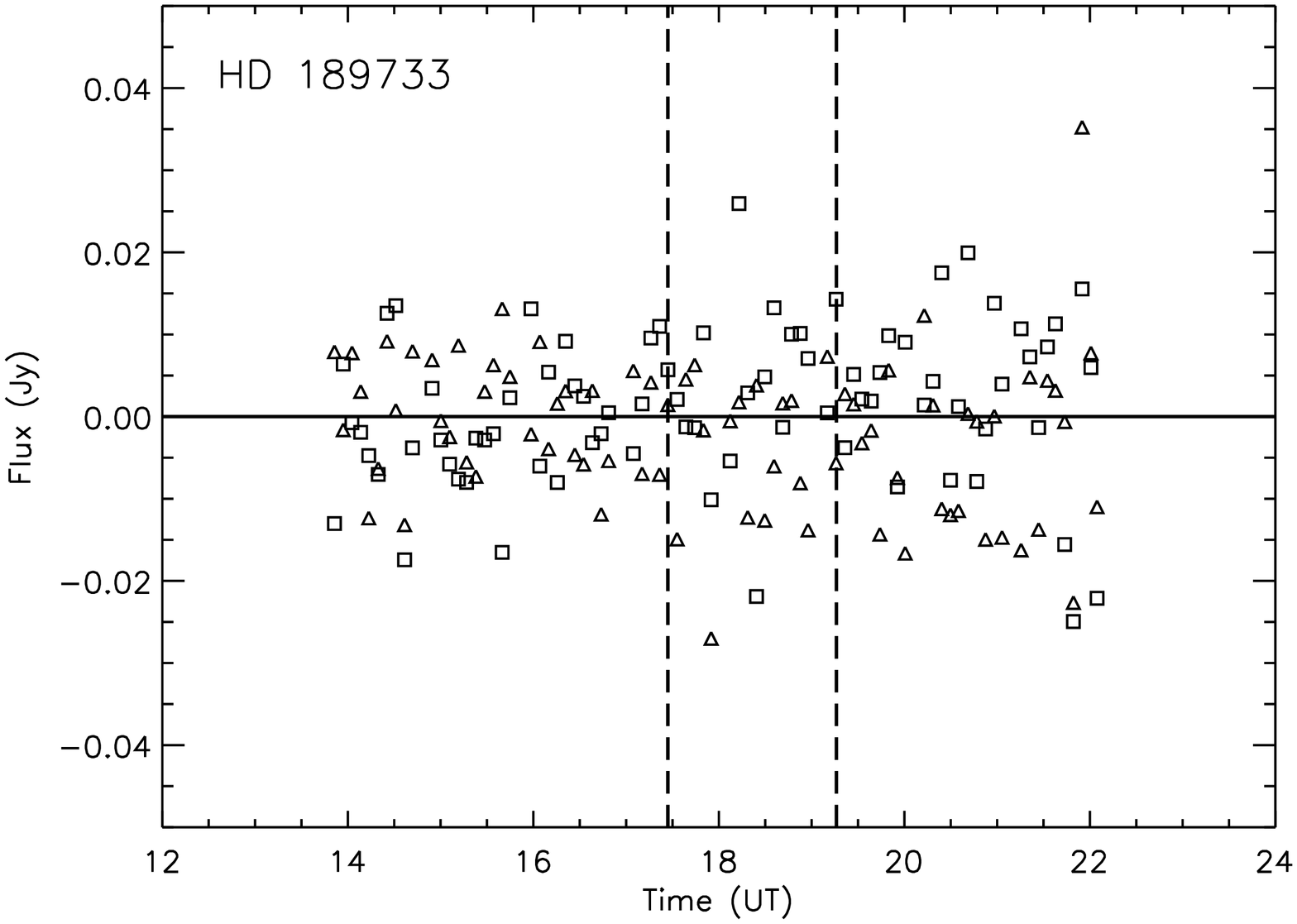}
}
\caption{Time series of the flux density in the direction of HD189733, for a sampling window 
of 339~seconds. The vertical dashed lines indicate the beginning and the end of the planet's
eclipse behind the host star. Triangles and squares correspond to the RR and LL polarizations, 
respectively. 
}
\label{HD189733_ts}
\end{center}
\end{figure}

\begin{figure}[tbh]
\begin{center}
\hbox{
\includegraphics[angle=90,width=\columnwidth]
{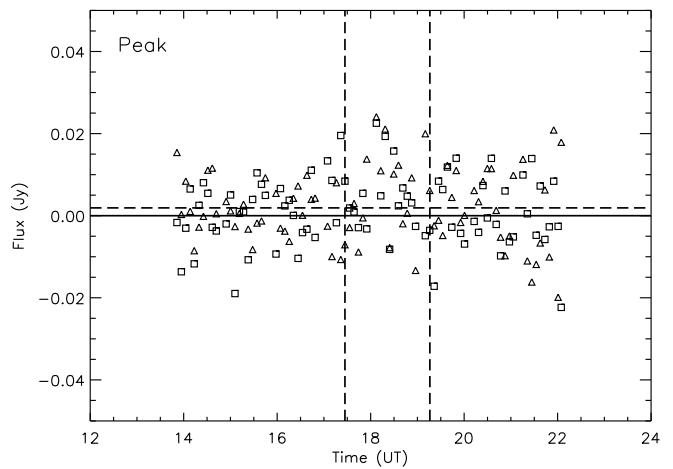}
}
\caption{Time series of the flux density in the direction 13\arcsec\ from HD189733 (see
text), with the same legend as in Figure~\ref{HD189733_ts}. The dashed horizontal line 
shows the flux level of this peak source at 1.9~mJy.}
\label{Peak_ts}
\end{center}
\end{figure}

While calibrating the visibilities, bad data points were flagged at various stages. 
The data for the antennas with relatively large errors in antenna-based gain 
solutions were examined and flagged over the corresponding time ranges. Some 
baselines were flagged, in view of large closure errors found on the bandpass calibrator. 
Channel- and time-based flagging of the data points corrupted by radio frequency 
interference (RFI) was made by applying a median filter with a $6\sigma$ threshold.
Cases of residual errors above $5\sigma$ were also flagged after a few rounds of 
imaging and self-calibration. The system temperature ($T_{sys}$) was 
found to vary among the antennas and also with 
the ambient temperature and elevation (Sirothia 
2009). In the absence of regular $T_{sys}$ measurements for GMRT antennas, 
this correction was estimated from the residuals of the calibrated data with 
respect to the model data. The corrections were then applied to the data. 
The final image was made after several rounds of phase self-calibration
until the change in S/N ratio between two successive rounds became 
less than 1\%. Lastly, we
applied one round of amplitude self-calibration, where the data were normalized 
using the median gain found for the entire data. 
The image was then corrected for the primary beam shape taken to be a 
Gaussian 
with a FWHM of 173.8\arcmin\ at 153~MHz.
For the central $8^\prime\times8^\prime$ regions shown in Fig.~\ref{HD209458_map}
and~\ref{HD189733_map}, $\sigma$ = 0.70\,mJy/beam for the HD189733 field and 
$\sigma$ = 1.0\,mJy/beam for the HD209458 field.

After completing the imaging, we generated the light curves for a synthesized-beam size region ($\sim$20\arcsec) centered on the coordinates 
of the two target stars. 
To obtain these light curves, we calculated model visibilities using the sources 
detected in the entire field-of-view excluding a synthesized beam-wide region centered 
on the target position ($\alpha_0, \delta_0$); we then subtracted this model 
from the final calibrated visibility data. 
The residual visibility data (RVD) were 
then phase-centered on $\alpha_0, \delta_0$ and can be averaged for any 
desired time bins to
generate the light curves. Figures~\ref{HD209458_ts} and~\ref{HD189733_ts} show 
the resulting light curves for the two targets.

From the light curves we can estimate a flux density
at the target position for any given time interval with its associated error bar. 
The flux density is estimated using a weighted mean of the data points in the light curves, 
where each measurement has a weight proportional to the number of visibilities used to obtain it. 
We checked that the light curves are free of correlated noise and that the RMS decreases with the square root
of the bin size. The error bars of the estimated flux densities are hence calculated as the
weighted standard deviation of the data points divided by the square root of the number of points
(for the 339 seconds bins we have a total of 160 and 176 data points in the two polarizations 
for HD189733 and HD209458, respectively). 
The resulting error bars on the estimated flux densities for the full integration time are found to be 
consistent with the independently estimated RMS 
of the final images: $\sim$1.0\,mJy for HD209458 and $\sim$0.7\,mJy for HD189733.

\section{Results}
\label{Summary of the results}

\subsection{HD209458b}

For the stellar+planetary emission, the GMRT 150~MHz image
of the HD209458 field shows no source at the target coordinates, setting
a 3$\sigma$ upper limit of about 3.6~mJy (Fig.~\ref{HD209458_map}).

Using the light curve, we also searched for the planet's 
emission at different orbital phases and any radio flares. 
With average flux densities of
$1.54$$\pm$$2.02$\,mJy, $-1.78$$\pm$$2.22$\,mJy, and $-1.39$$\pm$$2.03$\,mJy, 
respectively, before, during and after the secondary transit, no eclipse signature 
has been found.
Likewise, the running averages of the light curve, taken for time 
intervals between 1 and 30~minutes, revealed no flaring event.
In Fig.~\ref{HD209458_ts} we choose 339 seconds bin size to plot the light curve, 
this is a compromise between the apparent noise and the possibility 
to detect potential short lived emission flares.
We therefore conclude a non-detection of planetary radio emission. 
We note that the planet's eclipse times plotted in Fig.~\ref{HD209458_ts} 
(dashed vertical lines) 
are computed using the optical size of the planet. If the radio emission 
originates in a region larger than the planet, the ingress could happen earlier 
and the egress later. However, emission at frequencies around 150~MHz is expected to 
originate in the very inner regions of the planetary magnetosphere, 
and therefore the relevant ingress/egress times are expected to be very 
similar to those for the planet itself. 

\subsection{HD189733b}

The GMRT 150~MHz image in the HD189733 field shows no source at the target coordinates, 
setting a 3$\sigma$ upper limit of about 2.1~mJy (Fig.~\ref{HD189733_map}).
We also searched for the planet's 
emission at different orbital phases and for radio flares 
using the light curve (Fig.~\ref{HD189733_ts}). 
With average flux densities before, during, 
and after the secondary transit being,  
$-0.10$$\pm$$0.86$\,mJy, $-0.69$$\pm$$1.73$\,mJy, and $-0.46$$\pm$$1.51$\,mJy respectively,
no eclipse signature has been detected for this planet either.
Nonetheless, we report a marginal (2.7$\sigma$) detection of $\sim$1900$\pm$700~$\mu$Jy 
at $\alpha$=20h00m44.4s and $\delta$=+22\degr42\arcmin29.9\arcsec (J2000),
which is at 13\arcsec\ with a position angle of 50$\degr$ from the coordinates of HD189733.
The mean level in the light curve obtained in this direction
confirms the detection of an emission (Fig.~\ref{Peak_ts}). 
In the direction of this peak, the average flux densities before, during, and 
after the planetary eclipse are 
$1.01$$\pm$$0.86$\,mJy, $4.43$$\pm$$1.58$\,mJy, and $0.51$$\pm$$1.27$\,mJy, 
respectively, which is too noisy to find any eclipse signature.
We are consequently unable to reach a conclusion about 
the planetary or stellar origin of the 2.7$\sigma$  
signal detected at 150~MHz. 

The 13\arcsec\ shift between the measured position of the peak source and 
the absolute position of the star  
corresponds to about 250\,AU at the distance of the HD189733 planetary system. 
However, for a source detected at only a few sigma level, 
the position uncertainty is about the beam size. 
With a half-power beamwidth of 17\arcsec\ at the position angle of 50$\degr$, 
this shift is not significant and is well within the position uncertainty.
At the same time, the chance of this
emission being associated with a background/foreground source is fairly low.
Above the measured flux density of 1.9~mJy at 150~MHz we estimate a source density of 
about 0.3 per square arc minute in the target field, hence
the probability of such a chance projection within the
synthesized beam is just 5\%. 

\section{Discussion}
\label{Discussion}

Theoretical and observational aspects of the radio emission from 
extrasolar planets have been discussed in Zarka (2007), where
the generalized concept of flow-obstacle interaction was developed.
Accordingly, the clear non-detection toward HD\,209458 may be
understood because 
(1) the Earth was outside the planet's emission beam, 
at least at the time of observation, or (2) the emission is highly variable 
with flares lasting shorter than the temporal sampling achieved in our 
observations, or (3) the planetary emission was simply too weak 
intrinsically, or, perhaps more likely, (4) the planetary emission peaks at 
frequencies lower than 150~MHz because of the weakness of the planetary magnetic field. 

The first two scenarios (observer outside the emission beam, and short-duration 
flares) have been discussed in Lecavelier des Etangs et al.\ (2009). 
Although with a typical beaming
angle of $\sim$50\degr-60\degr\ the first scenario cannot be excluded,
both explanations for the non-detection are not the 
favored ones. Intrinsic weakness of the emission and/or its low-frequency 
appear to be more likely explanations for the non-detections. 

\subsection{Emission flux}

For the three mechanisms considered by Greissmeier (2007), i.e., kinetic 
emission, magnetic emission, and coronal mass ejection, the corresponding 
radio flux density estimates may reach 0.4, 900, and 30~mJy toward HD189733b, 
and 0.2, 150 and 10~mJy toward HD209458b, respectively. While all these
estimates are only suggestive, the magnetic emission estimate is the
highest. Unfortunately, this emission is predicted to peak at just a few MHz, 
which is much lower than the frequency range covered in the present observations.

Nonetheless, there is another possible scenario for the emission
mechanism.
For an intense magnetic field (of up to 40\,G in the case of HD189733; 
Moutou et al.\ 2007) the stellar magnetosphere can extend beyond the orbit of the  
extrasolar planets given the semi-major axis of just 0.03\,AU for HD189733b 
and 0.047\,AU for HD209458b
(see discussion in 
Jardine \& Cameron 2008). 
Assuming that the emission is produced by an
interaction between the planetary magnetosphere and the engulfing stellar corona, 
the use of Eq.~17 and the numerical values from the 
model of Jardine \& Cameron (2008), 
assuming a 10\% efficiency of conversion of the power of accelerated 
electrons into radio emission and taking an emission beam solid angle of 1.6\,sr,
the predicted radio flux densities are about 15\,mJy and 3\,mJy 
from HD189733\,b and HD\,209458\,b, respectively.  
With this model and assuming that
the present non-detection of HD189733\,b and HD209458\,b at 150\,MHz 
is caused by their low intrinsic radio luminosities, 
the upper limits of 2.1\,mJy toward HD189733b and 3.6\,mJy toward HD209458b 
can be translated into upper limits for the stellar coronal density. 
With a stellar magnetic field strength of 40\,G, 
we therefore obtain upper limits for the stellar coronal densities of 0.4 and 1.2~times the solar 
coronal density for HD189733 and HD209458, respectively. 
For HD209458 the stellar field strength, $B_*$, is not known, the upper limit for the coronal density
is smoothly dependent and scales with $B_*^{1/3}$.

\subsection{Emission frequency}

The principal mechanism advocated for radio emission is the electron-cyclotron maser radiation. 
It occurs at the local gyrofrequency 
$f_g= 2.8\left({B_p}/{\rm 1\,G}\right) {\rm MHz}$,
where $B_p$ is the planet's magnetic field strength. 
Our results could then be attributed to a weak planetary
magnetic field, such that the gyrofrequency falls below our observation 
frequency of 150~MHz. The corresponding planetary magnetic fields are
$<$50~G. Recall that the Jovian magnetic field estimated from the observed
spectral cut-off of its cyclotron radio emission is around 14~G. 

Fares et al. (2010) used their stellar magnetic field measurement
and its extrapolation to the distance of the planet 
to estimate the expected flux and frequency of the radio emission. 
They also estimated that the radio 
emission should be highly variable with a flux density in the range 7-220~mJy
(hence well within the detection limit of our observations), although 
with a frequency range of 0-6~MHz, possibly extending up to 20~MHz, if the model 
uncertainties are taken into account. Thus, our non-detections 
at 150 MHz could also be attributed to weak planetary magnetic field 
like the one assumed in the Fares et al. model. 

Finally, note that the cyclotron maser emission can also be quenched by  
too high a plasma frequency ($f_{pe}$) in the source region (Zarka 2007). 
With a 40~G magnetic field, the condition for quenching ($f_{pe}$ more 
than a tenth of the cyclotron frequency) implies an electron density higher 
than about 1.5$\times$10$^{6}$~cm$^{-3}$ in the low stellar corona, 
a condition that is not implausible.

\section{Conclusion}

In summary, the 150~MHz observations of HD189733\,b and HD209458\,b 
reported here represent an unprecedentedly deep search for planetary radio emission
at such low frequencies. 
The depth of these observations even surpasses some of the 
theoretical predictions published in the literature. 
The non-detections could be attributed to an inadequate 
temporal sampling rate of the observation, beam focusing, or 
intrinsic emission power being lower than the theoretical predictions.
Interestingly, the frequency of our observations would require planetary magnetic 
fields to be only a few times stronger than that of Jupiter, 
making weak planetary magnetic fields as a viable explanation for our non-detection.
We obtained a 1.9~mJy ($2.7\sigma$) detection of emission 
from a direction within a beamwidth of the position of HD189733, but the
light curve is not sensitive enough 
to reveal any correlation of the flux with the palnet's orbital phase,
thus precluding a definitive conclusion about the association of the radio emission
with the planet. While we continue our sensitive observations of nearby exoplanets 
using GMRT at 150 MHz,
similar searches at still lower frequencies are highly desirable and should
become feasible with UTR2 and LOFAR (in the short-term), 
and SKA (in the long-term).

\begin{acknowledgements}

We thank the staff of the GMRT for making these observations possible. 
GMRT is run by the National Centre for Radio Astrophysics (NCRA) 
of the Tata Institute of Fundamental Research (TIFR).
This program has been supported by the scientific award of the
"Fondation Simone et Cino Del Ducca". 
P.Z. activities in radio search for exoplanets are partly supported by
ANR program NT05-1\_42530 "Radio-Exopla".
\end{acknowledgements}

\end{document}